\documentclass[a4paper]{jpconf}
\usepackage{graphicx}
\begin{document}
\newcommand{\bea}{\begin{eqnarray}}
\newcommand{\eea}{\end{eqnarray}}
\title{ $\bar B_s\to K$ semileptonic decay from an Omn\`es improved
nonrelativistic 
quark model}

\author{C Albertus$^1$, E Hern\'andez$^2$, C Hidalgo-Duque$^3$ and J Nieves$^3$}

\address{$^1$ Departamento
de F\'\i sica At\'omica, Nuclear y Molecular  e Instituto Carlos
 I de F\'{\i}sica Te\'orica y Computacional, Universidad de Granada, Avenida
 de Fuentenueva s/n, E-18071
Granada, Spain}
\address{$^2$ Departamento
de F\'\i sica Fundamental e IUFFyM, Universidad de Salamanca, Plaza de la
Merced s/n, E-37008
Salamanca, Spain}
\address{$^3$ Instituto de F\'\i sica Corpuscular (IFIC), Centro Mixto
  CSIC-Universidad de Valencia, Institutos de Investigaci\'on de
  Paterna, Apartado 22085, E-46071 Valencia, Spain} 
\ead{albertus@ugr.es,gajatee@usal.es,carloshd@ific.uv.es,
jmnieves@ific.uv.es}

\begin{abstract}
We study the $f^+$ form factor for the $\bar B_s\to K^+\ell^-\bar\nu_\ell$
semileptonic decay in a nonrelativistic quark model. The valence quark 
contribution  is supplemented with a
$\bar B^*$-pole term that dominates the high $q^2$ region. 
To extend the quark model predictions from its
region of applicability near $q^2_{\rm max}=(M_{B_s}-M_K)^2$,
we use a multiply-subtracted
Omn\`es dispersion relation. We fit the subtraction constants to a combined
 input
from previous light cone sum rule results in the low $q^2$ region 
 and the  quark model results (valence plus $\bar B^*$-pole) in the high $q^2$
 region.   From this analysis, we obtain
$\Gamma(\bar B_s\to K^+\ell^-\bar\nu_\ell)=(5.47^{+0.54}_{-0.46})|V_{ub}|^2\times
10^{-9}\,{\rm MeV}$, which is about 10\% and 20\% higher than  predictions
based on Lattice QCD and  QCD light cone sum rules  respectively.
\end{abstract}

\section{Introduction}
 Playing  a critical
role in testing the consistency of the Standard Model of particle
physics and, in particular, the description of CP violation, $V_{ub}$ is still
the less well known element of the Cabibbo-Kobayashi-Maskawa (CKM) quark
 mixing matrix. Any new information that can be obtained from 
 experimentally unexplored reactions is thus relevant. 
 This is the case of the 
 $\bar B_s\to K^+ \ell^-\bar\nu_\ell$ semileptonic decay which is
expected to be observed at LHCb and Belle and that it could be used to obtain
an independent determination of $|V_{ub}|$.
In this contribution we present a study of this reaction. All the details
and further results to those presented here can be found in Ref.~\cite{albertus}.

The hadronic matrix element for the reaction can be parameterized in terms
of the $f^+(q^2)$ and  $f^0(q^2)$ form factors, of which only $f^+(q^2)$
plays a significant role for the case of a light lepton in the final 
state ($l=e,\mu$). In fact, for zero lepton masses, the differential 
decay width is given solely in terms of $f^+(q^2)$ as
\bea
\frac{d\Gamma}{dq^2}=\frac{G_F^2}{192\pi^3}\,|V_{ub}|^2\,
\frac{\lambda^{3/2}(q^2,M_{B_s}^2,M_K^2)}{M_{B_s}^3}\,|f^+(q^2)|^2
\label{eq:dgdq2}
\eea
with $G_F$  the Fermi decay constant and $\lambda$  the K\"allen function defined as 
$\lambda(a,b,c)=a^2 +b^2+c^2-2ab-2ac-2bc$.

\section{Results and discussion}
\begin{figure}[t]
\includegraphics[width=16pc]{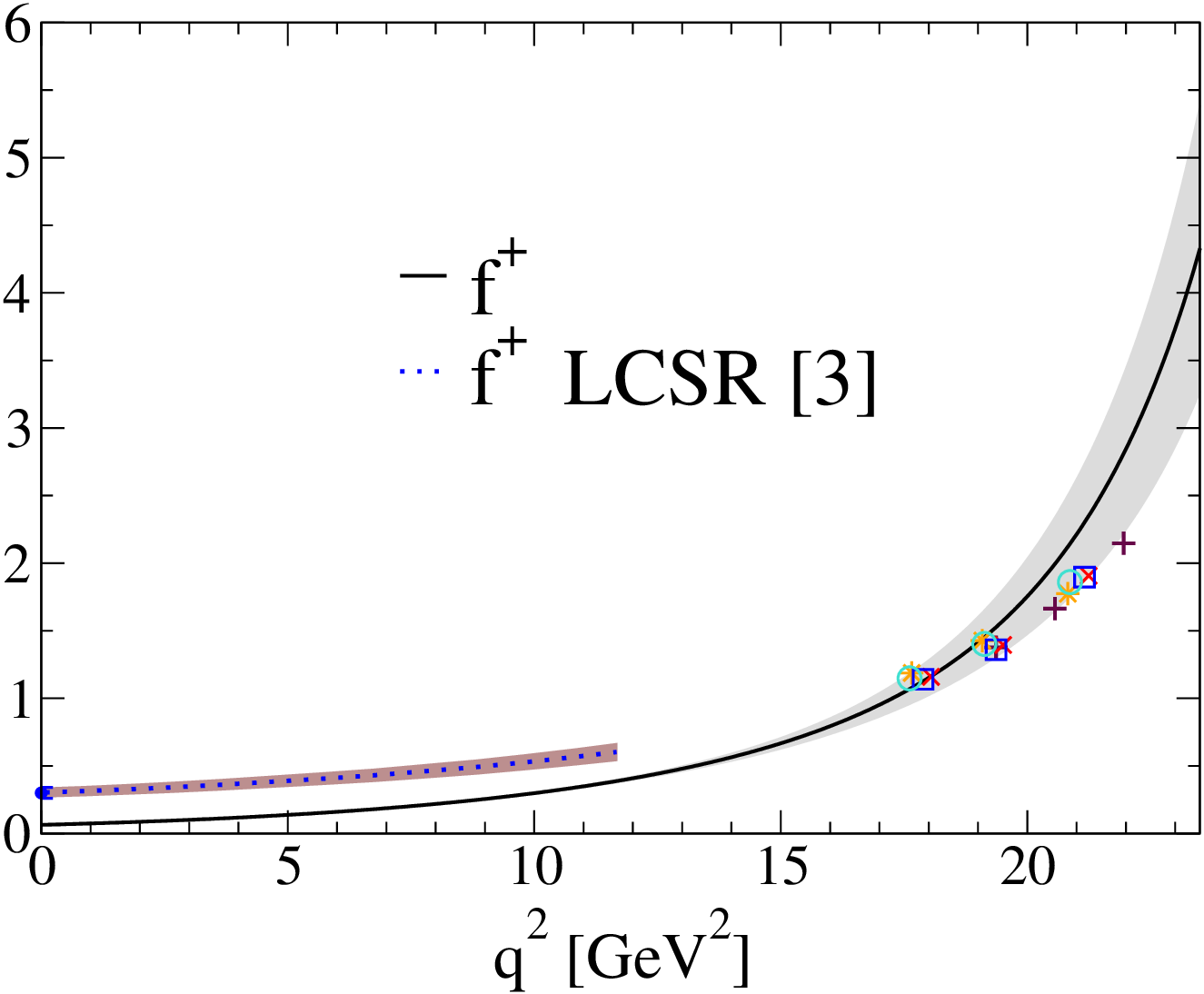}\hspace{2pc}%
\begin{minipage}[b]{20pc}\caption{\label{fig:total}$f^+(q^2)$  form factor evaluated in the quark
model (valence plus $\bar B^*$-pole
contributions). We also show the 
 results obtained in the LCSR calculation
of Ref.~\cite{Duplancic:2008tk} (dotted-line plus error band) 
and different lattice data  in the high $q^2$ region
reported in Ref.~\cite{Bouchard:2013zda}.}
\end{minipage}
\end{figure}

To obtain  the $f^+$ form factor we shall  follow  our
earlier work in Ref.~\cite{Albertus:2005ud}, where similar decays were analyzed,
and then we use the quark model to evaluate the valence plus $\bar B^*$-pole 
contributions to the form
factors. Calculational details can be found in \cite{albertus} and references therein.
Results are shown  in figure~\ref{fig:total}.
Taking into account theoretical uncertainties, shown as a band in the figure,
 we obtain  a reasonable description of the form factor in the high
$q^2$ region, as compared to 
 the preliminary lattice data recently reported in
Ref.~\cite{Bouchard:2013zda}. For high $q^2$, the $\bar B^*$-pole term 
dominates but the valence contribution  accounts for around 20\% of the total. 
However, there is a large
 discrepancy in the low $q^2$ region between the quark model 
and the light cone sum rule (LCSR) results obtained in Ref.~\cite{Duplancic:2008tk}.
Since the latter are reliable for low $q^2$, it is clear that the
non-relativistic quark model
does not provide a good reproduction of the form factor in that region of
$q^2$ where large relativistic effects are to be expected.

To obtain an $f^+(q^2)$ form factor valid for the whole $q^2$ region spanned by
the decay, we adopt the scheme in Refs.~\cite{Flynn:2006vr,
  Flynn:2007qd,Flynn:2007ii}, assuming  a
multiply subtracted Omn\`es functional ansatz that provides
a parameterization of the form factor constrained by unitarity and
analyticity properties. We take
\begin{eqnarray}
f^+(q^2)\approx\frac{1}{M^2_{B^*}-q^2}\prod_{j=0}^n\Big[f^+(q_j^2) 
\Big(M^2_{B^*}-q^2_j\Big)\Big]^{\alpha_j(q^2)}\ \ ,\ \ 
\alpha_j(q^2) = \prod_{j\ne k=0}^n \frac{q^2-q_K^2}{q^2_j-q^2_k}
\end{eqnarray}
for $q^2 < s_{\rm th}=(M_{B_s}+M_K)^2$ and where  $q_0, \cdots q_n^2 \in
]-\infty, s_{\rm th}[$ are the $(n+1)$ subtraction points. Note that despite
the factor $\frac{1}{M^2_{B^*}-q^2}$, the functional
form is not given by a single pole. The values of
 $f^+(q_j^2)$ are taken as free parameters and we fix them by making
 a combined 
fit to our quark model results in
the high $q^2$ region and to the LCSR results, taken from 
Ref.~\cite{Duplancic:2008tk}, in the low $q^2$ part. As in 
Ref.~\cite{Flynn:2007ii} we only use four subtraction points corresponding to
$q^2_j=0,\,q^2_{\rm max}/3,\,2q^2_{\rm max}/3,\,q^2_{\rm max}$.

\begin{figure}[t]
\includegraphics[width=18pc]{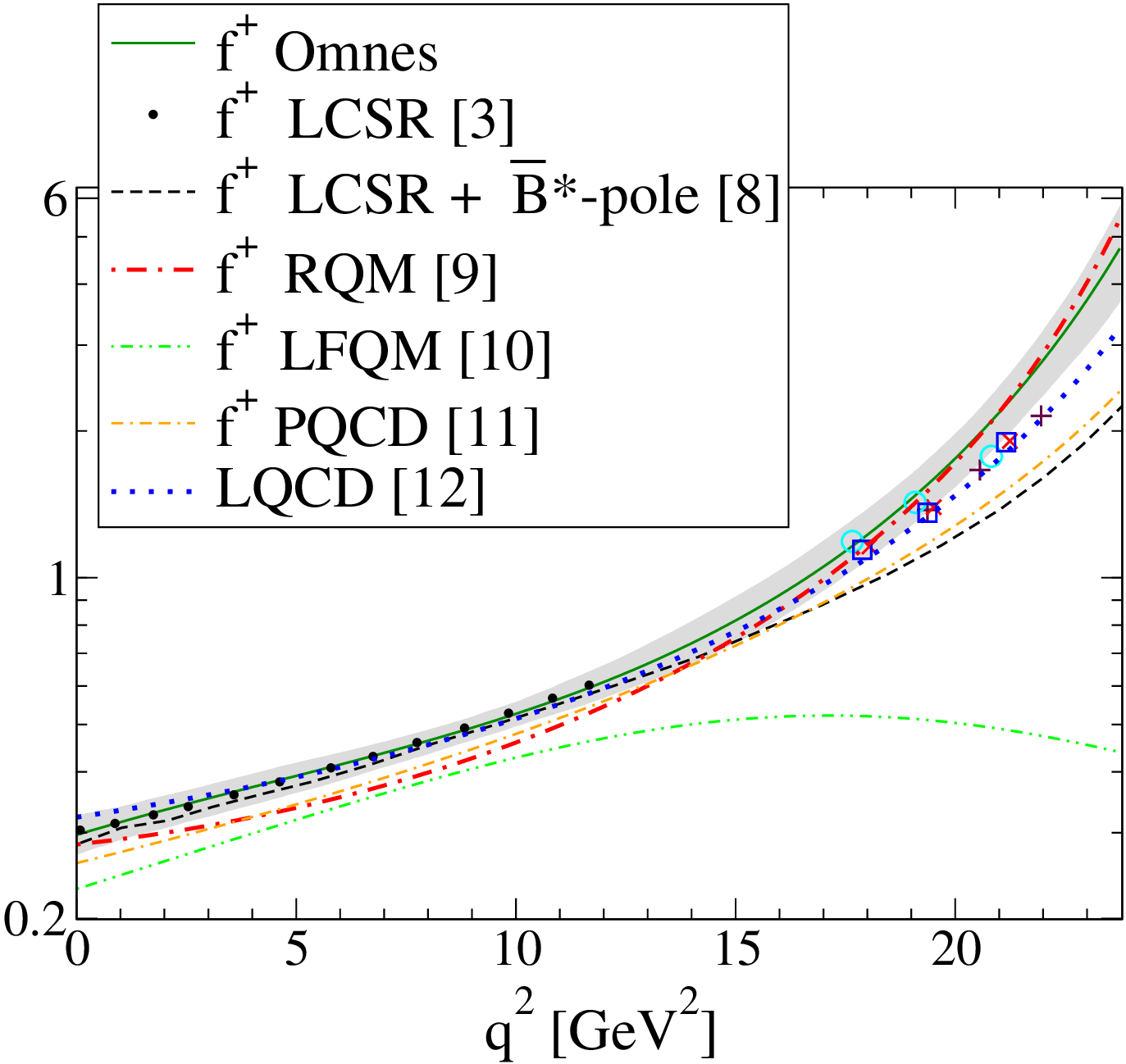}\hspace{2pc}%
\begin{minipage}[b]{18pc}
\caption{\label{fig:omnes2}Global comparison of
our final result (solid  line plus  68\% confidence level band) 
for the $f^+$ form factor with different calculations using
LCSR~\cite{Duplancic:2008tk}, LCSR$+\bar B^*$-pole fit~\cite{Li:2001yv}, 
RQM~\cite{Faustov:2013ima}, LFQM~\cite{Verma:2011yw}, 
PQCD~\cite{Wang:2012ab} and the  extrapolation to 
the physical region done
in Ref.~\cite{Bouchard:2014ypa} of the LQCD data from 
Ref.~\cite{Bouchard:2013zda} which is also
shown.}
\end{minipage}
\end{figure}

Our final result for
$f^+(q^2)$ together with its 68\% confidence level band is displayed in figure 
\ref{fig:omnes2}. There, we show a comparison
with different calculations using
LCSR~\cite{Duplancic:2008tk}, LCSR$+\bar B^*$-pole fit~\cite{Li:2001yv}, 
relativistic quark model (RQM)~\cite{Faustov:2013ima}, light front quark model 
(LFQM)~\cite{Verma:2011yw}, perturbative QCD (PQCD)~\cite{Wang:2012ab} and 
the  extrapolation to the physical region
done in Ref.~\cite{Bouchard:2014ypa} of the lattice  QCD (LQCD) results obtained in
Ref.~\cite{Bouchard:2013zda} (also
shown). In the LCSR calculation in 
Ref.~\cite{Duplancic:2008tk} the results are only given up to $q^2=10$\,GeV$^2$, 
whereas in 
Ref.~\cite{Verma:2011yw} no $\bar B^*$-pole 
contribution is included as can be seen by the behavior of the predicted form factor in
 the high $q^2$ region. All other  calculations  include the
$\bar B^*$-pole mechanism, but with different  strengths. 
 In Ref.~\cite{Faustov:2013ima}, where a RQM is used,  they obtain a form factor 
  similar to ours for high
$q^2$ values. However, their 
approach for 
 low and intermediate values of
$q^2$ should not be as appropriate as a LCSR one, which we include in our 
combined analysis. 
Calculations in Refs.~\cite{Wang:2012ab}  and \cite{Li:2001yv} give similar
results at high $q^2$ but the one in Ref.~\cite{Wang:2012ab} deviates from LCSR
evaluations at small $q^2$ values. The high $q^2$ results obtained in LQCD 
\cite{Bouchard:2013zda,Bouchard:2014ypa} 
 are in between the results obtained in the
 approaches of Refs.~\cite{Li:2001yv,Wang:2012ab} and the quark model ones 
 (both this work and the RQM
 calculation of Ref.~\cite{Faustov:2013ima}).  For very low $q^2$ however, the 
 central values of the LQCD
 extrapolation in Ref.~\cite{Bouchard:2014ypa} lie in the upper part of 
 the LCSR band. Our combined approach should be more adequate in that
 region  of $q^2$ since
  we use LCSR data
 to constraint our form factor.

\begin{figure}[t]
\includegraphics[width=18pc]{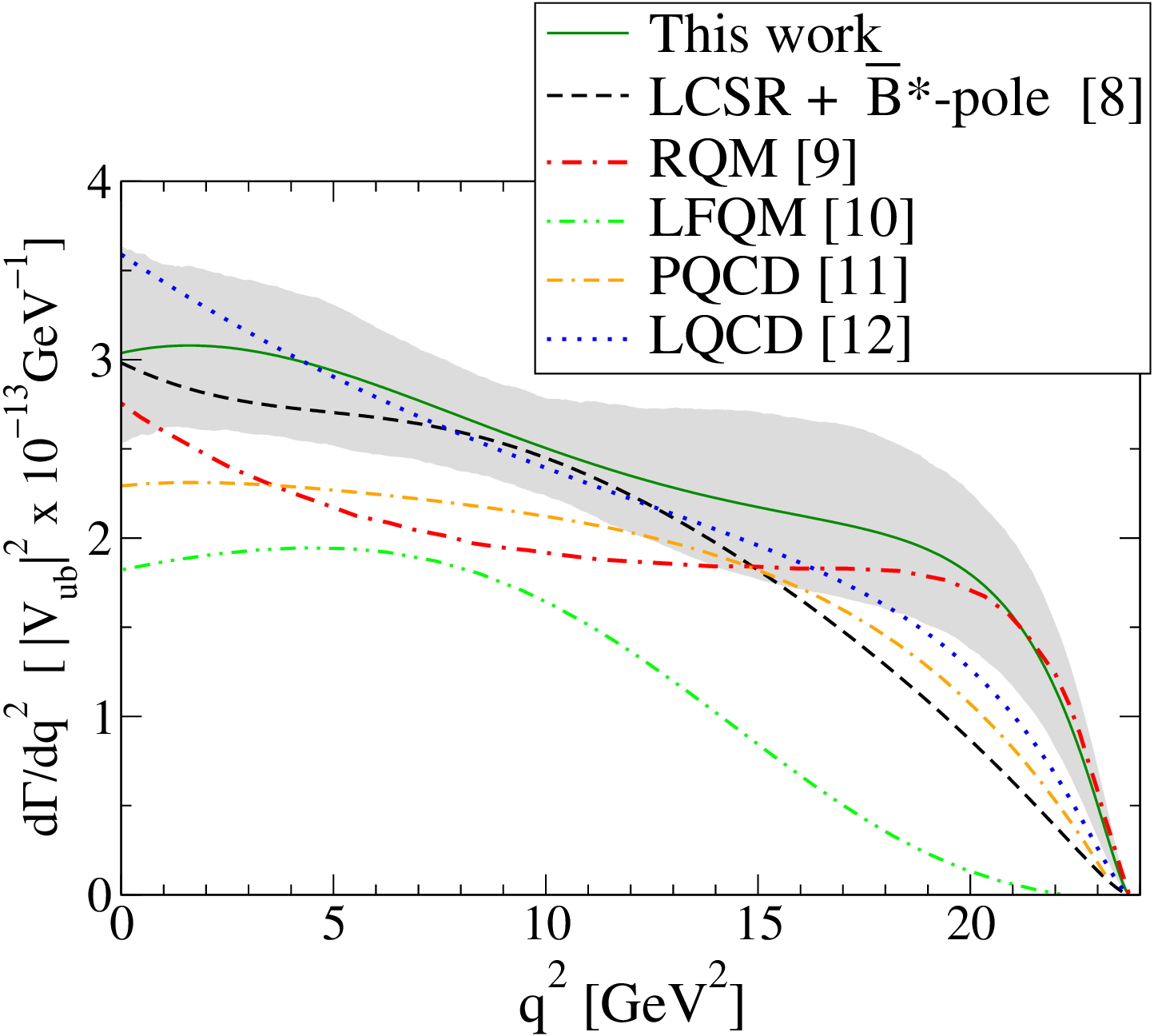}
\begin{minipage}[b]{18pc}
\caption{\label{fig:dgdq2}Differential decay width obtained in this work with the Omn\`es fit (solid line
plus 68\% confidence level band) and in LCSR$+\bar B^*$-pole fit~\cite{Li:2001yv}, 
RQM~\cite{Faustov:2013ima}, LFQM~\cite{Verma:2011yw} and
PQCD~\cite{Wang:2012ab} and LQCD~\cite{Bouchard:2014ypa} approaches.}
\end{minipage}
\end{figure}
The differential decay width,  together with
 its 68\% confidence level band, is displayed in  figure~\ref{fig:dgdq2}.  We
also show the differential decay width  from the calculations in 
Refs.~\cite{Li:2001yv,Faustov:2013ima,Verma:2011yw,Wang:2012ab,Bouchard:2014ypa}. 
For the integrated decay width we  obtain 
\bea
\Gamma(\bar B_s\to K^+\ell^-\bar\nu_\ell)=(5.47^{+0.54}_{-0.46})|V_{ub}|^2\times 10^{-9}\,{\rm
MeV} 
\eea
and a comparison with the results in other approaches is shown  in
table~\ref{tab:dw}.  The calculations in
Refs.~\cite{Li:2001yv,Faustov:2013ima} obtain results that are some 15\% smaller than
ours. The fact that their results are so similar when compared to each other
seems to be a coincidence. As seen in figure~\ref{fig:dgdq2}, their
differential decay widths deviate both for small and large $q^2$ values, but
those differences compensate in the integrated width. The
result of the PQCD calculation in Ref.~\cite{Wang:2012ab} is also
similar but with a larger uncertainty, around 50\%. The LFQM calculation in
Ref.~\cite{Verma:2011yw} gives a  much smaller result, in
part because  no $\bar B^*$-pole contribution seems to be included in that
approach. 
The LQCD  result in 
Ref.~\cite{Bouchard:2014ypa} is the one closest to ours. Its large 
uncertainty comes from the form factor
  extrapolation from high $q^2$, where the lattice points were obtained,
to the low $q^2$ region. Our result is the largest although 
we are compatible within uncertainties with the predictions of
Refs.~\cite{Li:2001yv,Faustov:2013ima,Wang:2012ab,Bouchard:2014ypa}.

\begin{table}[h]
\caption{\label{tab:dw}Decay width in units of $|V_{ub}|^2\times 10^{-9}\,{\rm MeV}$
  from several approaches. For the
result of Ref.~\cite{Li:2001yv} we have propagated a 10\% uncertainty in the
form factor. Results for Refs.~\cite{Faustov:2013ima,Verma:2011yw,Wang:2012ab}
have been adapted from Table IV in Ref.~\cite{Meissner:2013pba}.}
\begin{center}
\footnotesize{\begin{tabular}{l|cccccc}
\hline\hline

&This work&LCSR+$\bar B^*$-pole&RQM&LFQM&PQCD&LQCD\\
&&\cite{Li:2001yv}&\cite{Faustov:2013ima}&\cite{Verma:2011yw}&\cite{Wang:2012ab}
&\cite{Bouchard:2014ypa}\\
\hline
$\Gamma\ [\,|V_{ub}|^2\times 10^{-9}\,{\rm MeV}]$&$5.47^{+0.54}_{-0.46}$
&$4.63^{+0.97}_{-0.88}$&$4.50\pm0.45$&
$2.75\pm0.24$&$4.2\pm2.2$&$5.1\pm1.0$\\
\hline\hline
\end{tabular}}
\end{center}
\end{table}

\ack
 This research was supported by  the Spanish Ministerio de Econom\'{\i}a y 
 Competitividad and European FEDER funds
under Contracts Nos. FPA2010-21750-C02-02,  FIS2011-28853-C02-02,  
and the Spanish Consolider-Ingenio 2010 Programme CPAN (CSD2007-00042), by Generalitat
Valenciana under Contract No. PROMETEO/20090090, by Junta de Andalucia under
Contract No. FQM-225,
 by the EU HadronPhysics3 project, Grant Agreement
No. 283286,  and by the
University of Granada start-up Project for Young Researches contract No. PYR-2014-1.
C.A. wishes to acknowledge a CPAN postdoctoral contract and C.H.-D. thanks the support 
of the JAE-CSIC Program.


\section*{References}
\medskip
\begin{thebibliography}{9}
\bibitem{albertus} Albertus C, Hern\'andez E, Hidalgo-Duque C and Nieves 
 2014 $\bar B_s\to K$ semileptonic decay from an Omn\`es improved constituent 
quark model {\it Preprint  arXiv:1404.1001 [hep-ph]} 

\bibitem{Albertus:2005ud} 
  Albertus C, Flynn J M, Hern\'andez E, Nieves J and Verde-Velasco J M 2005
{\it  Phys.\ Rev.}\ D {\bf 72} 033002
  

\bibitem{Duplancic:2008tk}
  Duplancic G and Melic B 2008
 {\it Phys.\ Rev.\ } D {\bf 78}  054015


\bibitem{Bouchard:2013zda} 
  Bouchard C M, Lepage G P, Monahan C J, Na H and Shigemitsu J,
  $B$ and $B_s$ semileptonic decay form factors with NRQCD/HISQ quarks
  {\it Preprint arXiv:1310.3207 [hep-lat]}

\bibitem{Flynn:2006vr} 
  Flynn J M and Nieves J 2007 {\it Phys.\ Rev.}\ D {\bf 75} 013008

\bibitem{Flynn:2007qd} 
  Flynn J M and Nieves J 2007 {\it 
  Phys.\ Lett.}\ B {\bf 649} 269

\bibitem{Flynn:2007ii} 
  Flynn J M and Nieves J 2007 {\it Phys.\ Rev.}\ D {\bf 76} 031302


\bibitem{Li:2001yv} 
  Li Z H, Liang F Y, Wu X Y and Huang T 2001
  {\it Phys.\ Rev.}\ D {\bf 64} 057901

\bibitem{Faustov:2013ima} 
  Faustov R N and Galkin V O 2013
{\it  Phys.\ Rev.}\ D {\bf 87} 094028

\bibitem{Verma:2011yw}
 Verma R C 2012 {\it J.\ Phys.}\ G {\bf 39}  025005

\bibitem{Wang:2012ab} 
  Wang W F and Xiao Z J 2012 {\it Phys.\ Rev.}\ D {\bf 86}  114025

\bibitem{Bouchard:2014ypa} 
  Bouchard C M, Lepage G P, Monahan C, Na H and Shigemitsu J 2014 $B_s\to Kl\nu$
  form factors from lattice QCD
{\it Preprint arXiv:1406.2279 [hep-lat]}
  
\bibitem{Meissner:2013pba} 
  Mei\ss ner U G and Wang  W 2014 {\it JHEP} {\bf 1401} 107
  
\end{thebibliography}
\end{document}